\documentclass[pra,amsmath]{revtex4}
\usepackage[varg]{txfonts}   
\newcommand{\beq}{\begin{eqnarray}}
\newcommand{\eeq}{\end{eqnarray}}
\usepackage{esvect}
\usepackage{bbm}

\begin{document}

\title{Bringing Bell's theorem back to the domain of Particle Physics \& Cosmology}


\author{Beatrix C. Hiesmayr}

\affiliation{University of Vienna, Faculty of Physics, Boltzmanngasse 5, 1090 Vienna, Austria}
\email{Beatrix.Hiesmayr@univie.ac.at}

\begin{abstract}
John St. Bell was a physicist working most of his time at CERN and contributing intensively and sustainably to the development of Particle Physics and Collider Physics. As a hobby he worked on so-called ``\textit{foundations of quantum theory}'', that was that time very unpopular, even considered to be scientifically taboo. His 1964-theorem, showing that predictions of local realistic theories are different to those of quantum theory, initiated a new field in quantum physics: quantum information theory. The violation of Bell's theorem, for instance, is a necessary and sufficient criterion for generating a secure key for cryptography at two distant locations. This contribution shows how Bell's theorem can be brought to the realm of high energy physics and presents the first conclusive experimental feasible test for weakly decaying neutral mesons on the market. Strong experimental and theoretical limitations make a Bell test in weakly decaying systems such as mesons and hyperons very challenging, however, these systems show an unexpected and puzzling relation to another big open question: why is our Universe dominated by matter, why did the antimatter slip off the map? This long outstanding problem becomes a new perspective via the very idea behind quantum information.
\end{abstract}

\maketitle

\section{Introduction}

Only, since 2012 a promising proposal for testing Bell's 1964-theorem for systems usually produced at accelerator facilities, so called neutral $\mathcal{K}$-mesons, is on the market~\cite{Hiesmayr:2012}. Putting it to reality in current or upcoming accelerator facilities will be for several reasons challenging but feasible. However, from the theoretical point of view these systems at high energies are of great interest since --- as will be presented in this essay--- a puzzling relation between the information theoretic content and the violation of discrete symmetries was found~\cite{Hiesmayr1,Hiesmayr2,Hiesmayr3}. Discrete symmetries and their violation play an important role for the understanding of the forces and may have played or possible will play an important role in the development of our universe; in particular, why the antimatter did slip off.

This essay starts with an introduction to the cosmological question of the imbalance of matter and antimatter and proceeds by discussing discrete symmetries such as parity $\mathcal{P}$, charge-conjugation $\mathcal{C}$ and time reversal $\mathcal{T}$, their breaking and its implementations. After that Bell's theorem and in particular its application to quantum cryptographic protocols is reviewed. Herewith, all key ideas have been gather to relate the security of cryptographic protocols to the difference between a world of matter and antimatter. Last but not least half integer spin particles decaying weakly, so called hyperons, are discussed from an information theoretic perspective.

\section{How did the antimatter slip off the map of our Universe?}

The Russian mathematician Alexander Friedmann solved Einstein's equations of general relativity showing that these equations predict a continually expanding Universe. Since an expansion can not go on for ever sometime in the past it must have started out from a tiny spot, an explosion leading to our expanding Universe. In the aftermath of this Big Bang particle and antiparticle pairs were generated. As the Universe cooled, less and less particle-antiparticle pairs formed, while those already existing could have annihilated with each other to produce photons. If this picture is correct, there should be as many particles in the Universe as antiparticles.

Considering on the composition of an average cubic meter of our Universe we find $10^9$ photons, $1$ proton and no antiprotons. Tracing back to just after the Big Bang the same cubic metre should have had $10^9$ photons, $10^9$ antiprotons and $10^9+1$ protons. This sounds odd, why should there be slightly more baryonic matter particles than baryonic antimatter particles in our visible Universe? Where does this observed  imbalance of matter and antimatter come from?

There are two main obvious explanations for this asymmetry: either our Universe began with a small preference for matter, i.e. the total baryonic number of the universe is non-zero, or the universe was originally perfectly symmetric, but somehow a set of phenomena contributed to a small imbalance in favor of matter over time. This hypothetical physical processes that produces an asymmetry between baryons and antibaryons in the very early universe is dubbed baryogenesis. The second point of view is generally preferred, although there is no clear experimental evidence to favor one over the other one. Except the argument that an initial asymmetry should most likely have quickly eliminated due to thermodynamic considerations.

Short after 1964 the Russian nuclear physicist Andrei Dmitrievich Sakharov put forward three necessary conditions for the baryogenesis to occur, regardless of the exact mechanism~\footnote{In his paper ``\textit{Violation of $\mathcal{CP}$ invariance, $\mathcal{C}$, and baryon asymmetry of the universe}'' in JETP Letters 5, 24 (1967), re-published in Soviet Physics Uspekhi 34, 392 (1991), he did not explicitly list the three conditions.}. Firstly, obviously the baryon number $\mathcal{B}$ must have been violated in at least one process generating baryons and antibaryons. Thus there needs to exist processes of the type
\beq
a\longrightarrow b+c\;,
\eeq
where $a,c$  are particles with baryon number $\mathcal{B}=0$  and $b$ a baryon ($\mathcal{B}=1$). If the discrete symmetry charge-conjugation $\mathcal{C}$, which transforms a particle to its antiparticle or vice versa, is a symmetry of the universe, then the mirrored $\mathcal{C}$ reaction $\bar{a}\longrightarrow \bar{b}+\bar{c}$ has to have the same rate $\Gamma$, i.e.
\beq
\Gamma(a\longrightarrow b+c)=\Gamma(\bar{a}\longrightarrow \bar{b}+\bar{c})\;.
\eeq
Then even if the baryon number $\mathcal{B}$ is violated a conservation of charge-conjugation symmetry $\mathcal{C}$ does not produce an advantage of matter over antimatter for a long period of time. Thus one needs to require the violation of this symmetry as well. This is not quite enough, since if we consider a (hypothetical) $\mathcal{B}$ symmetry violating process $a\longrightarrow b_l+b_l$, which generates left-handed baryons, then if $\mathcal{CP}$ is a symmetry of Nature ($\mathcal{P}$\dots parity), the mirrored $\mathcal{CP}$ process  $\bar{a}\longrightarrow \bar{b}_r+\bar{b}_r$ ($r$\dots right handed) and hence we have
\beq
\Gamma(a\longrightarrow b_l+b_l)+\Gamma(a\longrightarrow b_r+b_r)=\Gamma(\bar{a}\longrightarrow \bar{b}_l+\bar{b}_l)+\Gamma(\bar{a}\longrightarrow \bar{b}_r+\bar{b}_r)\;.
\eeq
Thus, though the symmetry $\mathcal{C}$ leads to different rates, the combined symmetry $\mathcal{CP}$ acts to conserve the total baryon number $\mathcal{B}$. Indeed, before 1964 physicists assumed that antiparticles are just there to explain the unexpected violation of the parity $\mathcal{P}$ symmetry, discovered 1956, that was seen in the experiments of madame Wu (see next section~\ref{brokenmirrors}). Consequently, the symmetry $\mathcal{CP}$ needs to be violated as well such that more baryons are generated than antibaryons.

One striking fact is that the masses of particles and their antiparticles are identical. At the thermal equilibrium Boltzmann distribution law dictates any existing baryon asymmetry to turn back to the same amount. Hence there must have been a period, i.e. outside of the equilibrium, where different regions existed, some conserving, some violating the $\mathcal{B},\mathcal{C}$ and $\mathcal{CP}$ symmetries. After baryogenesis has taken place, the universe should have turned to equilibrium and no other process should have appeared reversing this asymmetry.

All of these conditions are compatible with today's observations, particulary with the predictions of the Standard Model. In particular we want to draw our attention in the next section to the peculiar violation of the $\mathcal{P}$ symmetry and the $\mathcal{CP}$ symmetry, found in the year 1956 and 1964, respectively.

\section{
Broken mirrors and the absolute definition of charges}\label{brokenmirrors}

Weak interactions, one out of four fundamental forces, are responsible for the decay of massive quarks and leptons into lighter quarks and leptons. When fundamental particles decay, it is very astonishing: we observe the particle vanishing and being replaced by two or more different particles. Although the total of mass and energy is conserved, some of the original particle's mass is converted into kinetic energy, and the resulting particles always have less mass than the original particle that decayed. The only matter around us that is stable is made up of the smallest quarks and leptons, which cannot decay any further. When a quark or lepton changes type (a strange quark changing to down quark, for instance) it is said to change flavor. All flavor changes are due to the weak interaction.

Until the mid-1950s, physicists thought that handedness, i.e. a reflection of right to left or vice versa, does not change subatomic process. Similar like a right-handed screwdriver would work equally well if they were manufactured to be left-handed, quantum physicists took for guaranteed that if a handed particle or process exists then its mirror-image should also exist. A left-handed particles would reflect via a left-right mirror transformation as a left-handed particle, but otherwise all other properties would happen in the same way and with the same rate. Also the behaviour would not essentially be altered by a mirror that transforms particles into antiparticles and vice versa.

Perhaps stimulated by the discovery of molecules that had due to their handedness very different behaviours Lee and Yang reexamined the evidence of the conservation of these two transformations, parity $\mathcal{P}$ and charge-conjugation $\mathcal{C}$. They found that it has not rigorously been proven. Hearing about that Madame Chein-Shiung Wu set up an experiment with radioactive cobalt in a strong magnetic field. If the symmetry $\mathcal{P}$ holds as many electrons from the $\beta$-decay should spray out in direction of the field as in opposite direction. Indeed, it was even not a small effect, a $40$ per cent asymmetry was seen! This was indeed overlooked since no one before Lee and Yang dared to question this discrete symmetry!

Then it was obvious to look also closer to the second discrete symmetry $\mathcal{C}$! Indeed the experiments revealed that also this particle-antiparticle symmetry was flawed.

Wolfgang Pauli, in a letter to Victor Weisskopf~\cite{Briefwechsel}, has written ``\textit{Now the first shock is over and I begin to collect myself again (as one says in
Munich).\dots It is good that I did not make a bet. I would have
resulted in a heavy loss of money (which I cannot afford); I did make a fool of
myself, however (which I think I can afford to do)— incidentally, only in letters
or orally and not in anything that was printed. But the others now have the
right to laugh at me. What shocks me is not the fact that ``God is just left-handed'' but the fact
that in spite of this He exhibits Himself as left/right symmetric when He expresses
Himself strongly. In short, the real problem now is why the strong interaction are
left/right symmetric. How can the strength of an interaction produce or create
symmetry groups, invariances or conservation laws? This question prompted
me to my premature and wrong prognosis. I don't know any good answer to
that question\dots}''

Somehow the idea established that the antiparticles come to rescue the broken parity symmetry, meaning that at least the combined symmetry $\mathcal{CP}$ should be unbroken. Getting familiar with the non-conservation of these discrete symmetries the long outstanding so called ``\textit{$\tau$-$\theta$ puzzle}'' found a solution. The neutral particles named $\tau$ and $\theta$ surprisingly having the same mass, however, decaying in strongly different ways with opposite parities turned out indeed to be the same particle, nowadays known by the name neutral $\mathcal{K}$-meson or kaon. Kaons are mesons that are composed by a quark and an antiquark, in particular the neutral kaon's quark content is a strange quark $s$ and a down quark $d$. The new quantum number, strangeness $S$, was introduced by Kazuhiko Nishijima, Tadao Nakano and Murray Gell-Mann in 1953, to explain the ``strange'' behaviour of these particles being produced quite often but decaying rather slowly: the strong force responsible in production process conserves the strangeness quantum number $S$, whereas the weak interaction is the one in charge for the subsequent decay which violates the strangeness conservation. The two different decay times with the two different decay channels, $2$ pions having $\mathcal{CP}=+1$  and $3$ pions having $\mathcal{CP}=-1$, of a single neutral $\mathcal{K}$-meson are thus naturally explained by the maximal violation of $\mathcal{C}$ and $\mathcal{P}$, respectively. In modern formalism one would say that a neutral kaon strangeness state $|\mathcal{K}^0\rangle$ is a superposition of a short-lived state $|\mathcal{K}_S\rangle$ (decaying into $2$ pions) and the long lived state $|\mathcal{K}_L\rangle$ (decaying into $3$ pions),
\beq
|\mathcal{K}^0\rangle &=&\frac{1}{\sqrt{2}} \left\lbrace |\mathcal{K}_S\rangle+|\mathcal{K}_L\rangle\right\rbrace
\eeq
and an antikaon state $|\overline{\mathcal{K}}^0\rangle$  is consequently defined by (where one puts the minus sign is not of physical importance)
\beq
|\overline{\mathcal{K}}^0\rangle &=&\frac{1}{\sqrt{2}} \left\lbrace -|\mathcal{K}_S\rangle+|\mathcal{K}_L\rangle\right\rbrace\;.
\eeq
Since both the kaon and its antimatter state decay into the same products, one cannot distinguish via the daughter particles the strangeness number, i.e. being a particle or an antiparticle. This means that if in a process a say antikaon is generated, it can oscillate in its particle state and vice versa, this is known under the term ``\textit{strangeness oscillations}'': $s\rightarrow \bar s\rightarrow s\rightarrow\dots$. Solving the Schr\"odinger equation for the two particle scenario one finds the following time evolution for an initial $|K^0\rangle$
\beq
|\mathcal{K}^0(t)\rangle &=&\frac{1}{\sqrt{2}} \left\lbrace e^{-\frac{\Gamma_S t}{2}}\cdot e^{-i m_S t}\, |\mathcal{K}_S\rangle+e^{-\frac{\Gamma_L t}{2}}\cdot e^{-i m_L t}\, |\mathcal{K}_L\rangle\right\rbrace\;,
\eeq
where $m_{S,L}$ and $\Gamma_{S,L}$ corresponds to the masses and decay constants of the short/long lived states. Note that between the decay rate $\Gamma_S$ and $\Gamma_L$ is a factor of about  $600$. So typically, a short lived state decays after $4cm$ and thus the long lived state decay peaks only after about $2.4m$! Was this combined symmetry $\mathcal{CP}$ really conserved or also flawed?

In 1963, James Cronin and Val Fitch set up an experiment at Brookhaven to have a closer look. Concentrating their data analyzes on the long lived kaons end of October 1963, they expected to see only $3$ pions, however, for one event out of thousand decays they did not find the third pion. Since obviously such a small effect could have many reasons, they put all their effort into the careful analyzes. Due to a formal mistake in the submission process they could not present their experiment at the Washington meeting in April 1964 giving them several more months to find the error. After trying everything to explain these $2$ pion decays they decided to go public. Indeed no error could be found, also the combined symmetry $\mathcal{CP}$ is broken, not maximally as the other ones, only by a tiny amount!

Consequently, the mass eigenstates become non-orthogonal, i.e.
\beq
\langle \mathcal{K}_S|\mathcal{K}_L\rangle=\frac{2 Re\{\varepsilon\}}{1+|\varepsilon|^2}\;,
\eeq
where $\varepsilon$ is the $\mathcal{CP}$ violating parameter and of the order $10^{-3}$.

\textbf{The electric charge gets absolutely defined by a neutral particle:} Because of $\mathcal{CP}$ violation, Nature allows for an ultimate way of defining the charge. Indeed, the electrically neutral kaon distinguishes positive charges from negative charges! E.g. in semileptonic decays ($\mathcal{K}^0\longrightarrow \pi^-+l^++\nu_l$ and $\overline{\mathcal{K}}^0\longrightarrow \pi^++l^-+\overline{\nu}_l$ with $l$ being an electron $e^-$ or a muon $\mu$) the kaon state decays slightly more often into a positive charged lepton than the antikaon state in a negative charged lepton state. Before meeting an alien from a distant world one certainly should find out whether they consist of antimatter or matter since shaking hands with antimatter would result in annihilation. Thus a kaon experiment is highly recommended before scheduling any meeting!

Considering symmetries of Nature is a powerful tool to explain empirical observations in physics, in particular in Particle Physics. There are three discrete symmetries: The spatial reflection that leads to parity $\mathcal{P}$ conservation if and only if there is no distinction between left and right. Discovering antiparticles another discrete symmetry, charge-conjugation $\mathcal{C}$, connects the partner with its antipartner and its conservation reflects the non-existence of an absolute electric charge. Last but not least there is the time reversal symmetry $\mathcal{T}$, stating that there should be no preferences for a forward or backward direction in time. So far, no violations of the combined symmetry $\mathcal{CPT}$ has been found, but experimental investigations are ongoing (see e.g. Refs.~\cite{KLOE1,KLOE2}). Since the combined symmetry $\mathcal{CP}$ has been found to be broken in meson-antimeson systems, also $\mathcal{T}$ symmetry violations have to exist to compensate such that the symmetry $\mathcal{CPT}$ is conserved (only recently violation of the $\mathcal{T}$ symmetry have been measured without referring indirectly to $\mathcal{CP}$ symmetry violations~\cite{DirectTime}). Now let us turn to quantum information theory to discuss how distant parties can generate a key for sending secret messages among them.

\section{Bell's theorem and the security of quantum cryptography}

John St. Bell was a physicist working at CERN and contributing intensively and sustainably to the development of Particle Physics and Collider Physics. His famous 1964-theorem, known nowadays as Bell's theorem, shows that predictions of local realistic theories are different to those of quantum theory. As, e.g. one application it has been found that the violation of Bell's theorem is a necessary and sufficient criterion for generating a secure key for cryptography at two distant locations, which we will show in some detail in the following.

In the famous Einstein-Podolsky-Rosen scenario a source produces two particles, which are separated and independently measured by Alice and Bob. Both parties can choose among two different measurement alternatives $i=n,n'$ for Alice and $j=m,m'$ for Bob. These settings yield either the outcomes $k,l=+1$  or $k,l=-1$.
Any classical or quantum correlation function can be defined, e.g., by
\begin{eqnarray}
E_{AB}(i,j)=\sum_{k,l} (k\cdot l)\; P_{AB}^{kl}(i,j)\;,
\end{eqnarray}
where $P_{AB}^{kl}(i,j)$ is the joint probability for Alice obtaining the outcome $k$ and Bob obtaining the outcome $l$, when they chose measurements $i$ and $j$, respectively. For local realistic theories Bell's locality assumption imposes a factorization of the joint probabilities. Bell inequalities are tests for correlations that can be simulated using only local resources and shared randomness (a modern terminology for local hidden variables) and have, therefore, at hitherto nothing to do with quantum theory. Inserting the probabilities derived by the rules of quantum mechanics, however, in some cases lead  to a violation of the inequality, i.e. to a contradiction between predictions of local hidden variable theories and quantum theory. For bipartite entangled particles with two degrees of freedom a tight Bell inequality is the famous Clauser-Horne-Shimony-Holt (CHSH) Bell inequality~\cite{CHSH}, i.e.
\begin{eqnarray}\label{chsh}
-2\leq S(n,m,n',m'):=E_{AB}(n,m)-E_{AB}(n,m')+E_{AB}(n',m)+E_{AB}(n',m')\leq 2\;.
\end{eqnarray}
In quantum mechanics the  $S(n,m,n',m')$-function is given by deriving the four quantum mechanical expectation values $E^{QM}_{AB}(n,m')(\rho)=Tr(O_n\otimes O_{m'}\rho)$ (where $O_i$ are appropriate operators and $\rho$ is the density matrix of the bipartite state). It is straightforward to prove that only entangled states can violate CHSH-Bell inequality, but not all entangled states violate the inequality. Moreover, the maximum violation $S^{QM}=2\sqrt{2}$ is reached for a maximally entangled state, e.g., the antisymmetric Bell state
\beq\label{antisymm}
|\psi^-\rangle&=&\frac{1}{\sqrt{2}}\biggl\lbrace |\Uparrow\rangle\otimes|\Downarrow\rangle-|\Downarrow\rangle\otimes|\Uparrow\rangle\biggr\rbrace\;.
\eeq

Let us extent the Einstein-Podolsky-Rosen scenario such that Alice and Bob measure randomly and independently one out of three specifically chosen observables each, where two of those observables are equal. Moreover, we assume that the source produces without loss of generality the maximally entangled antisymmetric Bell state~(\ref{antisymm}). Via a fully public open channel Alice and Bob announce their observable choices but not there measurement outcomes. Then we have two cases: Alice and Bob have chosen by chance the same or unequal observables. In case, Alice and Bob have chosen the same observable, since the source produces the antisymmetric Bell state, their measurement outcomes are perfectly anti-correlated and they can use these outcomes to both obtain a fully identical and random string of $``0''$ and $``1''$ (before, they have decided which measurement outcome is labeled $``0''$ and which one is labeled $``1''$). In the remaining case, they announce their outcomes (even in public) and use this data to compute the four quantum mechanical correlations functions of the CHSH-Bell inequality. If there is no eavesdropping, the CHSH-Bell inequality should be maximally violated. As proven in detail in Ref.~\cite{Ekert} any violation of Bell's theorem guarantees that an attack by an eavesdropper, even including the manipulation of the source, cannot reveal enough bit's of the string of the sifted key of Alice and Bob. Consequently, Alice and Bob can be sure based on the quantum laws that there generated key is secure! These correlations that violate Bell's theorem are stronger than anything that can be generated with classical physics!

Now we are prepared to connect both results, the existence of correlation stronger than those by classical physics and the violation of the discrete symmetries.

\section{Bell's theorem and the violation of the $\mathcal{CP}$ symmetry}

The main problems in testing Bell's theorem conclusively are limitations that arise from the experimental side. These are in particular that only the antisymmetric Bell state (compare with Eq.(\ref{antisymm})),
\beq
|\psi^-\rangle&=&\frac{1}{\sqrt{2}}\biggl\lbrace |\mathcal{K}^0\rangle\otimes|\overline{\mathcal{K}}^0\rangle-|\overline{\mathcal{K}}^0\rangle\otimes|\mathcal{K}^0\rangle\biggr\rbrace\;,
\eeq
is typically produced with high enough intensity and, secondly, only the strangeness content of neutral $\mathcal{K}$-mesons can be measured by an ``\textit{active}'' measurement procedure. Active measurements are a crucial requirement for any conclusive test of Bell's theorem since obviously if Alice and Bob have no control over their measurements it is straightforward to construct a local realistic theory resulting in the observed correlation. In particular, a decay event is a ``\textit{passive}'' measurement procedure, i.e. no experimenter has control over into which particles the meson will decay nor at which time this decay will occur. Though --as shown in reference~\cite{HiesmayrOpenQuantum}-- an decay can be viewed as an open quantum system, in particular modeled by a Markovian Lindblad master equation, the decay states have to be included. This, in particular, means that one is not allowed to normalize to only surviving pairs. These are all subtle points that need to be taken into account for testing Bell's theorem conclusively in the domain of high energy systems. The requirement of an ``\textit{active}'' measurement procedure rules out all other meson system due to short decay constants, except the neutral $\mathcal{K}$-meson system. The second requirement that all information available has to be considered makes it hard to find a Bell inequality that is violated for the observed constants in the $\mathcal{K}$-meson system.

\textbf{Actively measuring the strangeness content of neutral $\mathcal{K}$-mesons:} The experimenter places at a certain distance from the source a piece of matter that forces the incoming neutral $\mathcal{K}$-meson beam to interact with the material and to reveal the strangeness content, i.e. being at that distance in the state $|\mathcal{K}^0\rangle$ or in the state  $|\overline{\mathcal{K}}^0\rangle$. Since Bell's theorem tests against all local realistic theories one is not allowed to ignore the fact that the neutral kaon could have decayed before. Therefore the question that one has to raise has to include that information, i.e. one has to ask: ``\textit{Are you at a certain distance from the source in the state $|\overline{\mathcal{K}}^0\rangle$ or not?}'', which is obviously different to the question ``\textit{Are you at a certain distance from the source in the state $|\overline{\mathcal{K}}^0\rangle$ or in the state $|\mathcal{K}^0\rangle$?}''.

To test Bell's theorem given by the $S$-function, Eq.(\ref{chsh}), one has to compute four expectation values for such active measurements of strangeness given for the antisymmetric Bell state for different distances (that one can always convert in proper times since the velocity for a given experimental setup is known). Surprisingly, an optimization over all possible distances (times) does not show any value higher than $2$ and $-2$, i.e. no contradiction to local realistic theories. Why is this the case?

The point is that the oscillation in comparison to the two decay constants is too slow or, equivalently, the decay is too fast in comparison to the oscillation. Since we cannot obviously change the natural constants of elementary particles, we have to search for a different Bell inequality. Unfortunately, the CHSH-Bell version is already the most tight one. In reference~\cite{Hiesmayr:2012} the authors derived a new type of Bell's inequality for decaying system by including the decay property into the derivations of the bounds from local realistic theories. They assumed that any local realistic theory must also describe the well experimentally tested time evolution of single mesons correctly. This is the more striking since in a typical accelerator experiment $\mathcal{K}$-mesons are only generated in pairs, in huge contrast to typical photon experiments. Observing only a single event on one side, one knows with very high probability that the other one existed but due to purely experimental reasons was not detected.

With this new Bell inequality~\cite{Hiesmayr:2012} taking the decay property into account without spoiling the conclusiveness, the authors show which time regions have to be investigated experimentally to reveal correlations that are stronger than those allowed by classical physics. Surprisingly, though investigating strangeness oscillation the $\mathcal{CP}$ symmetry violation plays a crucial role! Asking the question ``\textit{Are you at a certain distance from the source in the state $|\mathcal{K}^0\rangle$ or not?}'' or ``\textit{Are you at a certain distance from the source in the state $|\overline{\mathcal{K}}^0\rangle$ or not?}'' makes the difference, i.e. leading in one case to a violation in the other one not!

\textbf{Consequently, the security of cryptography protocols depends in a given setup on analyzing the particle or the antiparticle content! How odd is Nature!}

\section{Bell's theorem and the violation of the $\mathcal{P}$ symmetry}

In the last section we discussed neutral mesons which are spinless particles. Hyperons are half-integer spin particles that are baryons containing in addition to up or down quarks also one or more strange quarks. They decay via the weak interaction violating the $\mathcal{P}$ symmetry. The Standard Model of elementary particles predicts also tiny contribution of  $\mathcal{CP}$ violating processes, however, no violation of the $\mathcal{CP}$ symmetry has been up to now experimentally found. In this section we discuss the quantum information theoretic content of weakly decaying hyperons and discuss whether Bell's theorem can be tested for these weakly decaying systems.

Any closed quantum system's dynamic is given by the Schr\"odinger equation, i.e. a unitary evolution. Some times one is only interested in a part of the closed quantum system or has only access to a part of the system, for instance a spin in a heat bath. The dynamics of the system of interest, the open quantum system, can be derived by the unitary evolution of the total system, system of interest plus environment, and taking the partial trace of the environmental degrees of freedom (for an introduction to open quantum systems consult e.g. Ref.~\cite{BreuerPetruccione,RivasHuelga}). On the other hand, if the total Hamiltonian is not known, one can study the dynamics of open quantum systems by a proper parametrization of the dynamical map. Any kind of time evolution of a quantum state $\rho$ can always be written in the form~\cite{Salgado}
\beq\label{Lindbladequation}
\rho(t)& =&
\sum_i \mathcal{K}_i(t,t_0,\rho(t_0))\;\rho(t_0)\;\mathcal{K}_i^\dagger(t,t_0,\rho(t_0))
\eeq
where the operators $\mathcal{K}_i$ are in general dependent on the initial time $t_0$ and state $\rho(t_0)$ and are often called the Kraus operators. In particular, the dynamical map defines a universal dynamical map if it is independent of the state it acts upon. This is only the case if and only if the map is induced from an extended system with the initial condition $\sigma_{\rm{total}}(t_0)=\rho(t_0)\otimes \rho_{\rm{environment}}(t_0)$ where $\rho_{\rm{environment}}(t_0)$ is fixed for any $\rho(t_0)$. This is exactly the above described scenario.


In Ref.~\cite{HiesmayrHyperon} it has been shown  that any hyperon decay process can be modeled efficiently by an open quantum formalism, i.e. via Kraus operators. Typically, the directions of the momentum of the daughter particles of a hyperon are measured. This distribution is connected to the initial spin state of the hyperon. In the weakly decay process there are two interfering amplitudes, one conserves and one violates the parity $\mathcal{P}$ symmetry. The momentum distribution computes to ($\theta,\phi$ are the angular coordinates of the momentum direction of one daughter particle and $\rho_{spin}$ the density operator corresponding to the spin degrees of freedom of the decaying hyperon)
\beq
I(\theta,\phi)&=& Tr_{spin}(K_+\;\rho_{spin}\; K_++K_-\;\rho_{spin}\; K_-)\nonumber
\eeq
where the Kraus operators have the conceptually simple form ($\omega_\pm>0$)
\beq
K_\pm&=\sqrt{\omega_{\pm}}\; |\vv{\omega}_1\pm\vv{\omega}_2\rangle\langle \vv{\omega}_1\pm\vv{\omega}_2|\,:=\, \sqrt{\omega_{\pm}}\; \Pi_{\vv{\omega}_1\pm\vv{\omega}_2}\nonumber
\eeq
with $\omega_++\omega_-=1$. The two Blochvectors $\vv{\omega}_{1,2}$ have to be orthogonal, $\vv{\omega}_{1}\cdot\vv{\omega}_{2}=0$, since the transition is completely positive and are chosen such that they have maximal length $|\vv{\omega}_1\pm\vv{\omega}_2|^2=s (2s+1)$ ($s$\dots spin number).

A Blochvector expansion of a density matrix is generally given by $\rho=\frac{1}{d}\{\mathbbm{1}_d+\vv{b}\cdot\vv{\Gamma}\}$ where $d$ is the dimension of the system~\cite{KrammerBertlmann}. Since we are dealing with spin-degrees of freedom we have $d=2s+1$ and we can choose as a set of orthonormal basis the generalized Hermitian and traceless Gell-Mann matrices  $\vv{\Gamma}$  (for $s=\frac{1}{2}$ they correspond to the Pauli matrices). Given this structure we can reinterpret the weak decay process as an incomplete spin measurement of the decaying particle
\beq\label{ErgInt}
I(\theta,\phi)&=&\omega_+\; Tr (\Pi_{\vv{\omega}_1+\vv{\omega}_2}\,\rho_{spin})+\omega_-\; Tr (\Pi_{\vv{\omega}_1-\vv{\omega}_2}\,\rho_{spin})\nonumber\\
&=&\frac{1}{(2s+1)}\left\{1+\left(\vv{\omega_1}+(\omega_+-\omega_-)\,\vv{\omega_2}\right)\cdot\vv{s}\right\}
\eeq
where $\vv{s}$ is the Bloch vector representation of $\rho_{spin}$, i.e. $\vv{s}=Tr(\vv{\Gamma} \rho_{spin})$.
With probability $\omega_+$ the spin state of the hyperon is projected onto direction $\vv{\omega}_1+\vv{\omega}_2$ or with the remaining probability $\omega_-$ the initial spin state is measured along the direction $\vv{\omega}_1-\vv{\omega}_2$. Thus the weak process can be associated to a spin measurement with an imperfect Stern-Gerlach apparatus (switching with probability $\omega_{\pm}$ the magnetic field). The imperfection has two causes: Firstly, the difference $(\omega_+-\omega_-)$ equals an asymmetry (denoted in the following by $\alpha$) and is a typical measurable constant for each hyperon. The asymmetry corresponds to interference contrast (visibility) times the cosine of the phase shift of the two interfering amplitudes, one is parity $\mathcal{P}$ conserving and one violates the symmetry $\mathcal{P}$. Secondly, the two directions $\vv{\omega}_1\pm\vv{\omega}_{2}$ are typical for the spin number $s$. Indeed, for $s=\frac{1}{2}$ the Blochvector $\vv{\omega}_1$ is zero, thus only two directions, $\pm\vv{\omega}_{2}$, are chosen by Nature.

Entangled hyperons can be produced, e.g. by proton-antiproton annihilations. The introduced open quantum formalism allows for a straightforward extension by the tensor product of the Kraus operators~\cite{HiesmayrHyperon}. Let us assume that (i) there is no initial correlation between the momentum degrees of freedom and the spin degrees of freedom and (ii) there is no entanglement between the momentum degrees of freedom. Experiments~\cite{PS185Exp1,PS185Exp2,PS185Exp3}, e.g. for the spin-$\frac{1}{2}$ $\Lambda$ hyperon and $\bar\Lambda$ anti-hyperon, suggest that the initial spin state is a maximally entangled Bell state (except for backward scattering angles). Therefore without loss of generality we can assume that (iii) the spin degrees of freedom of the particle and antiparticle are produced in the antisymmetric Bell state (compare with Eq.(\ref{antisymm}))
 \beq
|\psi^-\rangle&=&\frac{1}{\sqrt{2}}\biggl\lbrace |\Uparrow_\Lambda\rangle\otimes|\Downarrow_{\bar\Lambda}\rangle-|\Downarrow_{\bar\Lambda}\rangle\otimes|\Uparrow_{\Lambda}\rangle\biggr\rbrace\;.
\eeq
 Then the computation of the angular distribution of the momenta of the two daughter particles of the $\Lambda$ and $\bar\Lambda$ results in
\beq
I(\theta_\Lambda,\phi_\Lambda;\theta_{\bar\Lambda},\phi_{\bar\Lambda})&=& \frac{1}{4}\left\lbrace 1-\alpha_\Lambda \alpha_{\bar \Lambda}\; \vv{n}_\Lambda \cdot\vv{n}_{\bar\Lambda}\right\rbrace\;.\nonumber\\
\eeq
Since the Bloch vectors
\beq
n_{\Lambda/\bar\Lambda}=\left(\begin{array}{c}\sin\theta_{\Lambda/\bar\Lambda}\cos\phi_{\Lambda/\bar\Lambda}\\
\sin\theta_{\Lambda/\bar\Lambda}\sin\phi_{\Lambda/\bar\Lambda}\\
\cos\theta_{\Lambda/\bar\Lambda}\end{array}\right)
\eeq
are multiplied, $\vv{n}_\Lambda\cdot\vv{n}_{\bar\Lambda}$, by the constants $\alpha_\Lambda\cdot\alpha_{\bar\Lambda}$, T\"ornqvist~\cite{Tornqvist} concluded that the hyperon $\Lambda$ decays ``\textit{as if it had a polarization $\alpha_\Lambda$ tagged in the direction of the $\pi^+$ (coming from the $\bar\Lambda$) and vice versa}''. The knowledge of how one of the $\Lambda's$ decayed --or shall decay (since time ordering is not relevant)-- reveals the polarization of the second $\Lambda$. He concludes that this is the well-known Einstein-Podolsky-Rosen scenario.\\
\\
\textbf{Does the imperfection of the spin measurement allow for detection of entanglement?}\\
\\
In general entanglement is detected by a certain observable that can witness the entanglement content, i.e. a Hermitian operator $\mathcal{W}$ for which holds $Tr(\mathcal{W}\rho)<0$ for at least one state $\rho$ and $Tr(\mathcal{W}\rho_{sep})\geq 0$ for all separable states $\rho_{sep}$. For the antisymmetric Bell state such an optical entanglement witness is given by $\mathcal{W}=
\frac{1}{3}(\mathbbm{1}\otimes\mathbbm{1}+\sum_i \sigma_i\otimes\sigma_i)
$ (any other witness can be obtained by local unitary transformations). Since the weak interaction only allows for an imperfect spin measurement we have to multiply the spin part by $\alpha_\Lambda \alpha_{\bar \Lambda}$. Thus the entanglement witness for the $\Lambda\bar\Lambda$ system results in
\beq\frac{1}{3}-\alpha_\Lambda \alpha_{\bar \Lambda}\geq 0\;\forall\;\rho_{sep}\;,\eeq
which is clearly violated since $\alpha_\Lambda \alpha_{\bar \Lambda}=0.46\pm0.06$~\cite{ParticleDataBook}. Therefore, the measurement of the correlation functions $\langle\sigma_i\otimes\sigma_i\rangle$ in $x,x$ and $y,y$ and $z,z$ directions of the $\Lambda$ and $\bar\Lambda$ reveals entanglement. Let us here emphasize that a re-normalization (dividing by $\alpha_\Lambda \alpha_{\bar \Lambda}$) is not proper since also an mixed separable state may give the value up to $\frac{1}{3}$. Generally, one can say that the asymmetries lead to imperfect spin measurements which shrink the observable space. Equivalently, we can say that the given interferometric device leads to a shrinking of the Hilbert space of the accessible spin states.\\
\\
\textbf{However, does the imperfection of the spin measurement allow for detection of correlations stronger than those of classical physics?}\\
\\
For that we have to investigate Bell's inequalities and in principle all its variants. The CHSH-Bell type one, Eq.~(\ref{chsh}), leads to~\cite{HiesmayrHyperon}
\beq
\alpha_\Lambda\alpha_{\bar\Lambda}\stackrel{\textrm{for all local realistic theories}}{\leq}\frac{1}{\sqrt{2}}\;.\eeq
This is clearly not violated since $\alpha_\Lambda\alpha_{\bar\Lambda}\approx (0.46\pm0.06)$! However, here we anyway missed a requirement for any conclusive test of Bell's theorem: \textit{active} measurements! The weak interaction chooses the quantization axes $\pm\vv{\omega}_2$ spontaneously, we just know the probabilistically which one (with the probabilities $\omega_{\pm}$). Thus Bell's theorem cannot be tested in this way!

\section{Outlook}

We discussed Bell's theorem in the realm of high energy physics. We have seen that a conclusive version that can be experimentally be put to reality is very involved, however, both mesons and hyperons offer a new theoretical perspective since their properties are connected to violations of the discrete symmetry parity $\mathcal{P}$ and the combined discrete symmetry charge-conjugation-parity $\mathcal{CP}$. This surprising relation between Bell's theorem and discrete symmetry breaking, in turn, can be attributed to the unsolved problem of why we live in a universe dominated by matter, a problem which itself links to distinct and fundamental questions in physics. This discovery may pave the way to provide a first understanding for a more fundamental rule in Nature and to whether information theoretic considerations played a key role in the development of our universe (or may play).\\
\\
\textbf{Acknowledgements:}
The author gratefully acknowledges the Austrian Science Fund projects FWF-P26783 and wishes to thank the COST Action MP1006 ``Fundamental Problems in Quantum Physics''.

%
%
%

\end{document}